\documentclass[aps,prl,floatfix,twocolumn,reprint,amsmath,amssymb,superscriptaddress,UTF8]{revtex4-2}
\usepackage{amsfonts}
\usepackage{mathrsfs}
\usepackage{amsmath}
\usepackage{color}
\usepackage{natbib}
\usepackage{graphicx}
\usepackage{bm}
\usepackage{amssymb}
\usepackage{xspace}
\usepackage{epstopdf}
\usepackage{dcolumn}
\usepackage{longtable}
\usepackage{array}
\usepackage{makecell}
\usepackage{tabulary}
\usepackage{multirow}
\newcolumntype{C}[1]{>{\centering\arraybackslash}p{#1}}
\newcolumntype{L}[1]{>{\flushleft\arraybackslash}p{#1}}

\usepackage{tcolorbox} 
\newtcolorbox{highlightbox}{colback=red!10!white, colframe=red!50!black, arc=10pt, boxrule=1.5pt}  

\usepackage{multirow}
\usepackage{epsfig}
\usepackage[normalem]{ulem}
\usepackage{amsmath}
\usepackage{braket}
\usepackage{bbold}
\usepackage{natbib}

\usepackage[colorlinks=true, letterpaper=true, pdfstartview=FitV, linkcolor=blue, citecolor=blue, urlcolor=blue]{hyperref}

\makeatletter

\newcommand{\Rmnum}[1]{\expandafter\@slowromancap\romannumeral #1@}
\makeatother

\begin{document}

\title{Nonvolatile Electrical Control of Spin via Sliding Fractional Quantum Multiferroics}

\author{Jiajun Lu}
\affiliation{Key Lab of advanced optoelectronic quantum architecture and measurement (MOE), Beijing Key Lab of Nanophotonics $\&$ Ultrafine Optoelectronic Systems, and School of Physics, Beijing Institute of Technology, Beijing 100081, China}
\affiliation{International Center for Quantum Materials, Beijing Institute of Technology, Zhuhai, 519000, China}

\author{Mu Tian}
\affiliation{Key Lab of advanced optoelectronic quantum architecture and measurement (MOE), Beijing Key Lab of Nanophotonics $\&$ Ultrafine Optoelectronic Systems, and School of Physics, Beijing Institute of Technology, Beijing 100081, China}
\affiliation{International Center for Quantum Materials, Beijing Institute of Technology, Zhuhai, 519000, China}

\author{Chaoxi Cui}
\affiliation{Key Lab of advanced optoelectronic quantum architecture and measurement (MOE), Beijing Key Lab of Nanophotonics $\&$ Ultrafine Optoelectronic Systems, and School of Physics, Beijing Institute of Technology, Beijing 100081, China}
\affiliation{International Center for Quantum Materials, Beijing Institute of Technology, Zhuhai, 519000, China}

\author{Zhi-Ming Yu}
\affiliation{Key Lab of advanced optoelectronic quantum architecture and measurement (MOE), Beijing Key Lab of Nanophotonics $\&$ Ultrafine Optoelectronic Systems, and School of Physics, Beijing Institute of Technology, Beijing 100081, China}
\affiliation{International Center for Quantum Materials, Beijing Institute of Technology, Zhuhai, 519000, China}

\author{Run-Wu Zhang}
\email{zhangrunwu@bit.edu.cn}
\affiliation{Key Lab of advanced optoelectronic quantum architecture and measurement (MOE), Beijing Key Lab of Nanophotonics $\&$ Ultrafine Optoelectronic Systems, and School of Physics, Beijing Institute of Technology, Beijing 100081, China}
\affiliation{International Center for Quantum Materials, Beijing Institute of Technology, Zhuhai, 519000, China}

\author{Yugui Yao}
\affiliation{Key Lab of advanced optoelectronic quantum architecture and measurement (MOE), Beijing Key Lab of Nanophotonics $\&$ Ultrafine Optoelectronic Systems, and School of Physics, Beijing Institute of Technology, Beijing 100081, China}
\affiliation{International Center for Quantum Materials, Beijing Institute of Technology, Zhuhai, 519000, China}
\date{\today}
\begin{abstract}
\hyphenpenalty=750 
We propose a fractionally quantized polarization induced by interlayer sliding in bilayer altermagnets, unveiling a previously unrecognized multiferroic phase termed sliding fractional quantum multiferroicity (SFQM). This unconventional magnetic phase uniquely integrates sliding ferroelectricity with fractional quantum ferroelectricity, enabling highly efficient switching and nonvolatile electrical control of spin.~Unlike conventional multiferroics, SFQM simultaneously exhibits lattice-scale atomic displacements, ultralow switching barriers, and spin splitting, giving rise to a large fractionally quantized polarization and strong magnetoelectric coupling. Through symmetry analysis and first-principles calculations, we identify bilayer altermagnet Ca(CoN)$_2$ and its family materials as promising candidates hosting SFQM. In contrast to gate-controlled schemes, the spin-layer coupling in SFQM is intrinsically induced by spontaneous electrical and layer polarization, requiring no sustained gate field and exhibiting nonvolatile character. This mechanism enables nonvolatile electrical control of spin through biaxial sliding, where displacements along the \textit{x}- and \textit{y}-axes generate opposite polarization directions in the layer-dependent electrical polarization. Furthermore, SFQM exhibits a fully switchable anomalous Hall effect and a pronounced magneto-optical response, which can be utilized for its detection and distinction. These findings highlight the promising role of sliding-mediated couplings among unconventional magnetism, fractional quantum ferroelectricity, and stacking order in realizing electrically controllable two-dimensional multiferroics.
\end{abstract}
\maketitle

\begin{figure}[htb]
	\begin{center}
	\includegraphics[width=\columnwidth]{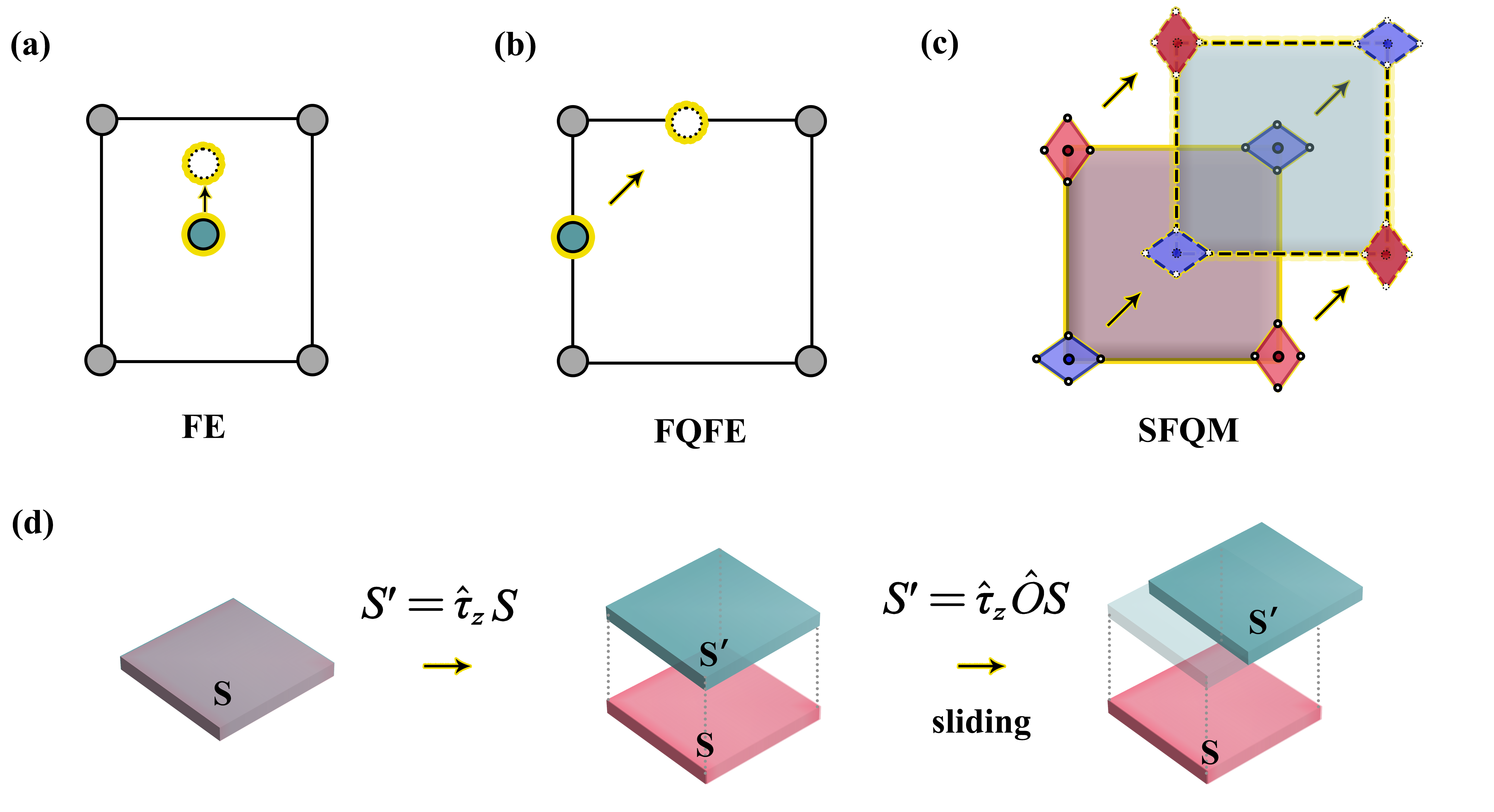}
	\caption{Concept of sliding fractional quantum multiferroicity (SFQM). (a),(b) Comparison of conventional ferroelectricity (FE) and the proposed fractional quantum ferroelectricity (FQFE).  (c) In altermagnets, sliding induces fractionally quantized polarization for a SFQM state.~Grey, green, red, and blue spheres denote ligand ions, movable nonmagnetic ions, and magnetic ion clusters with opposite spins, respectively.~(d) Underlying structure and symmetry of the altermagnetic bilayer. The top layer $S'$  is obtained by applying an operation to the bottom layer $S$. Here, $\hat{\tau}_z=$\{$E|\tau_z$\} is a trivial out-of-plane translation operator which does not change the symmetry of the bilayer, and $\hat{O}=\{E|\tau_{o}\}$ is the sliding operator, where $\tau_{o}$ is the in-plane sliding part. 
	}
	\label{fig:one}
	\end{center}
\end{figure}

\textit{\textcolor{blue}{Introduction}}--
Coupling between coexisting ferroic orders, such as ferroelectricity \cite{FE1,FE2,FE3,FE4}, ferromagnetism \cite{FM1,FM2,FM3}, and ferroelasticity \cite{Fel1,Fel2,Fel3}, enables unique device functionalities and underpins the surge of interest in multiferroicity within condensed matter physics and materials science \cite{MF1,MF2,MF3}.~As one of the cornerstone ferroic orders, conventional ferroelectricity (FE) stems from small atomic displacements that give rise to a spontaneous polarization [Fig.~\ref{fig:one}{\color{blue} (a)}]. It has held a lasting fascination since its initial discovery in perovskite oxides \cite{FE3,FE4}, serving as a prototype for exploring emergent physical mechanisms and enabling a broad spectrum of device functionalities. 

Among existing ferroelectric research frameworks, two prominent themes stand out:~the energy barrier and the electric polarization.~For the former, recent advances in sliding ferroelectric  mechanisms  \cite{hy-li-2017,hy1,hy2-WTe2-bilayer,hy3,hy4,hy5} have demonstrated a pathway toward achieving ultralow switching barriers while maintaining nonvolatile polarization states. Regarding the latter, emergent  fractional quantum ferroelectricity (FQFE) \cite{FQFE-NC,FQFE-PRL}, which is characterized by its ability to induce polarization even in nonpolar systems and distinguished by a fractionally quantized polarization component, represents a unique class of quantum ferroelectric states, as shown in Fig.~\ref{fig:one}{\color{blue} (b)}. These complementary breakthroughs invite a transformative possibility: could the integration of sliding ferroelectricity with fractionally quantized polarization defines a new paradigm in multiferroics, thereby pushing the understanding and applications of quantum ferroic materials beyond the fundamental limits of traditional ferroelectrics?

A paradigm shift is currently reshaping our understanding of magnetic orders, which constitutes another pivotal element of multiferroics. 
Conventionally, magnetic materials have been classified into two fundamental categories: ferromagnets and antiferromagnets. This binary framework, however, imposes a critical trade-off in multiferroic applications, forcing a choice between a strong net magnetization for effective magnetoelectric coupling and complete immunity to external stray fields. 
The recent emergence of altermagnetism \cite{PRX1,PRX2,AM-o-1,AM-o-2,AM-o-3,AM-o-4,AM-o-5,AM-o-6,AM-o-7} as a distinct third collinear magnetic phase fundamentally reconfigures this long-standing dichotomy. Altermagnets uniquely combine momentum-dependent spin-split electronic bands with zero net magnetization, thereby offering an unprecedented platform for efficient magnetoelectric coupling. 

More recently, leveraging the intrinsic attributes of altermagnetism within multiferroic systems has opened new frontiers in cross-coupling, facilitating bidirectional conversion between electrical means and magnetic states. These rapid developments naturally lead to a compelling question: could the synergistic integration of altermagnetism with emerging ferroelectric mechanisms, particularly sliding ferroelectricity and fractional quantum ferroelectricity, gives rise to a fundamentally new class of multiferroics that transcends existing material limitations?

In this work, we propose a unified framework from these advances by introducing sliding fractional quantum multiferroicity (SFQM), defined as a multiferroic phase that coherently combines sliding ferroelectricity, fractional quantum ferroelectricity, and altermagnetism, as illustrated in Fig.~\ref{fig:one}{\color{blue}(c)}. This SFQM paradigm uniquely merges the ultralow switching barrier of sliding ferroelectrics, the fractionally quantized polarization of FQFE, and the momentum-locked spin splitting of altermagnets within a unified platform. Based on symmetry analysis and first-principles calculations, we predict the bilayer altermagnet Ca(CoN)$_2$ and its family materials as promising candidates for realizing SFQM, where the cooperative interplay of three ferroic orders is symmetry-allowed. In SFQM, spin manipulation is enabled by sliding means rather than an external gate field, ensuring nonvolatile operation without a sustained gate field. Sliding along orthogonal crystal directions switches the polarization states, allowing for effective and nonvolatile control of the spin-layer coupling.
Furthermore, we demonstrate that SFQM hosts a fully switchable anomalous Hall effect and exhibits pronounced magneto-optical responses.~Our work establishes a new design principle for quantum multiferroics and provides a concrete pathway toward multifunctional spintronic devices by harnessing sliding-mediated couplings among unconventional magnetism \cite{UM1,UM2}, fractional quantum ferroelectricity \cite{FQFE-NC,FQFE-PRL}, and stacking order \cite{BS1,BS2}.

\textit{\textcolor{blue}{General analysis for SFQM}}--
To elucidate the physics of SFQM, we begin with a square-lattice model. By generalizing the concept of mobile atoms for FQFE to sliding van der Waals layers and coupling it with the electronic structure of a bilayer altermagnet, we propose a previously unrecognized multiferroic phase---the SFQM phase. This realization is schematically shown in Fig.~\ref{fig:one}{\color{blue}(c)}.

Having introduced the SFQM phase, we now present a general procedure for its realization based on symmetry analysis. The approach consists of the following steps:
(i) Starting from a van der Waals altermagnetic monolayer ($S$), we apply the stacking operator $\hat{\tau}_z$ to obtain $S'_{H}=\hat{\tau}_zS$. 
The resulting bilayer $B_{H} = S + \hat{\tau}_zS$ exhibits AA stacking, corresponding to the high-symmetry intermediate phase $H$ of the SFQM state. Here, $\hat{\tau}_z=\{ E|\tau_z\}$ denotes an out-of-plane translation that preserves bilayer symmetry.
(ii) By fixing one layer and sliding the other by a fractional multiple of the lattice vector, we arrive at the $L_1$ phase. Formally, $S'_{L_1}=\hat{\tau}_z\hat{O}_{L_1}S$, where $ \hat{O}=\{E|\tau_{\alpha}\}$ and $\tau_{\alpha}$ represent the sliding displacement along the $\alpha$ direction.
(iii) Additional SFQM phase such as $L_2$ can be constructed via $L_{2}=\mathcal{O}L_{1}$, where $\mathcal{O}$ is a symmetry operation present in $H$ but not in $L_1$. Crucially, $L_2$ must be accessible from $H$ via interlayer sliding.
(iv) Finally, we verify whether the transition from $L_1$ to $L_2$ qualifies as SFQM by examining if the sliding displacement is fractionally quantized.
\begin{figure}[htb]
	\begin{center}
	\includegraphics[width=\columnwidth]{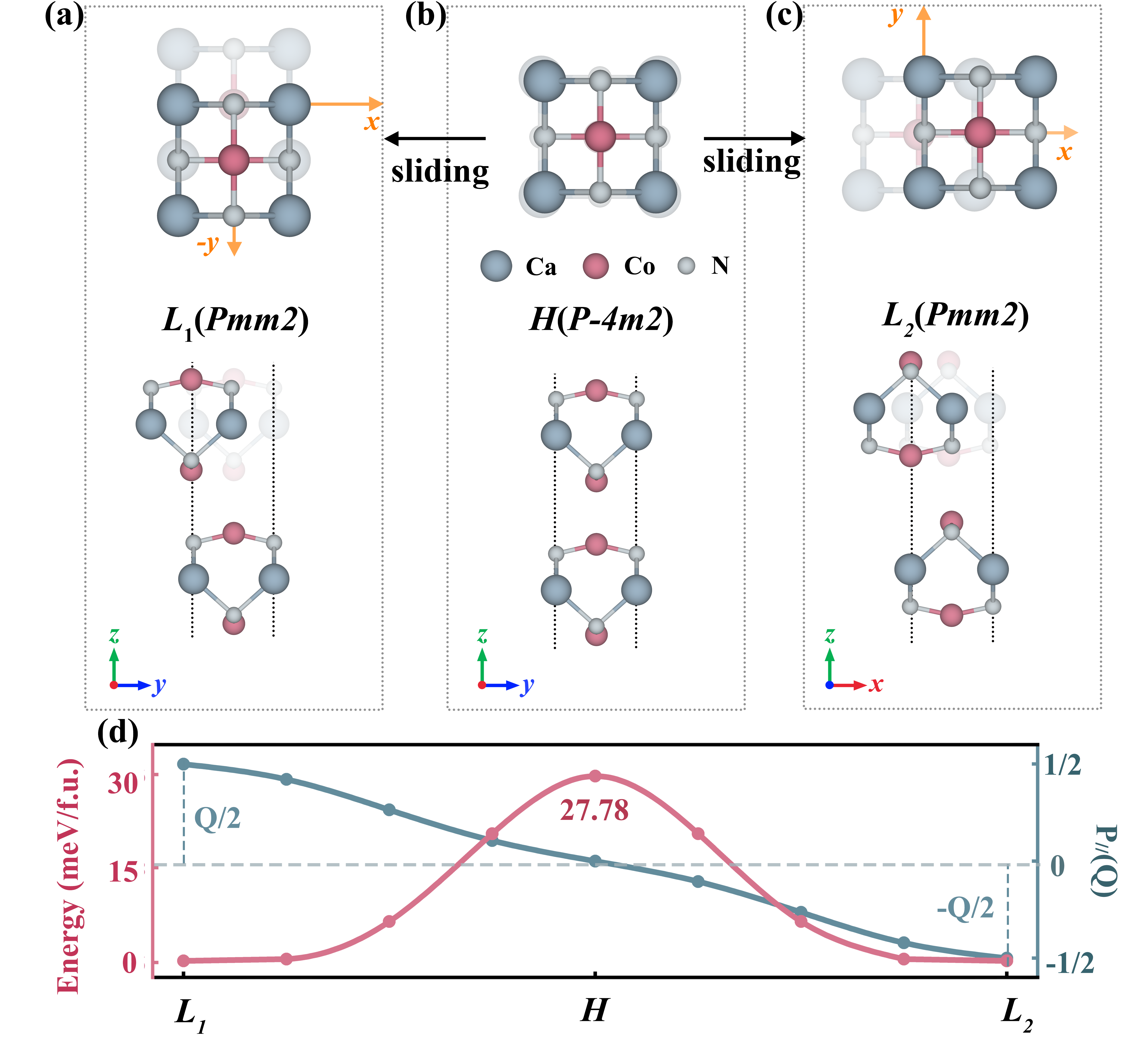}
	\caption{Sliding pathway and fractional polarization in bilayer Ca(CoN)$_2$. Top and side views of the bilayer structure in the (a) $L_1$, (b) $H$, and (c) $L_2$ phases, with the in-plane sliding displacement of $\frac{1}{2}\mathbf{a} + \frac{1}{2}\mathbf{b}$ indicated. (d) The phase transition pathway characterized by the energy barrier from nudged elastic band (NEB) calculations and the evolution of the polarization. The primitive cells for the $L_1$, $H$, and $L_2$ phases are depicted, showing the fractional polarizations $|\mathbf{P}_1| = |\mathbf{P}_2| = \frac{1}{2}\mathbf{Q}$.
	}
	\label{fig:two}
	\end{center}
\end{figure}

\textit{\textcolor{blue}{Material candidates}}--
Layered altermagnets are highly advantageous for realizing SFQM in practical devices, owing to their superior scalability, tunability, and integration compatibility over bulk materials. This potential is bolstered by the advent of an increasingly diverse family \cite{CaCoN} of atomically thin altermagnets, synthesized chemically or exfoliated mechanically, following pivotal advances in stabilizing long-range magnetic order in the two-dimensional limit. Among these, the recently proposed Ca(CoN)$_2$ family emerge as promising candidates. Owing to its inherent layered structure and altermagnetic ground state, it is rendered an ideal platform for exploring the proposed SFQM phenomenon.

Bulk Ca(CoN)$_2$ (mp-1443952) crystallizes in an orthorhombic lattice with an AA-stacked layered structure, facilitating the exfoliation of bilayers.
We construct the bilayer system by defining the bottom and top layers as $S$ and $S'$, respectively. The high-symmetry AA-stacked phase ($H$) is obtained by an out-of-plane translation: $S'_H=\hat{\tau_z}S$, where $\hat{\tau_z}=\{E|\tau_z\}$ is a trivial out-of-plane translation operator which does not change the symmetry of the bilayer, yielding $B_H=S+S'_H$, as shown in Fig.~\ref{fig:two}{\color{blue}(b)}.
To realize the polar $L_1$ phase, we fix the bottom layer and slide the top layer in-plane. This is described by $S'_{L_{1}} = \hat{\tau_z} \hat{O}_{L_{1}}S$, where $\hat{O}_{L_{1}}=\{E|\tau_{-y}\}$, and $\tau_{-y}$ represents a translation of  half unit cell along the $-y$ direction [Fig.~\ref{fig:two}{\color{blue}(a)}].
The $L_2$ phase can be described by  $S'_{L_2} = \hat{\tau}_z \hat{O}_{L_{2}} S$, where $\hat{O}_{L_{2}}=\{E|\tau_{x}\}$, and $\tau_{x}$ represents a translation of half unit cell along the $x$ direction [Fig.~\ref{fig:two}{\color{blue}(c)}].

A key aspect of SFQM is the polarization quantum $\mathbf{Q} = (e/\Omega) \mathbf{a}_{[110]}$, where $e$, $\Omega$, and $\mathbf{a}_{[110]}$ are the electron charge, unit-cell volume, and lattice vector along [110], respectively. For bilayer Ca(CoN)$_2$, we find $ |\mathbf{Q}| \approx 5.0\ e\mathrm{\AA}$ per unit cell. The computed in-plane polarizations for the $L_1$ and $L_2$ phases are $+2.5\, e\mathrm{\AA}$ and $-2.5\, e\mathrm{\AA}$ per unit cell, respectively [Fig.~\ref{fig:two}{\color{blue}(d)}], confirming that $\mathbf{P} = \pm \frac{1}{2} \mathbf{Q}$. The resulting polarization difference $\Delta \mathbf{P} = \mathbf{Q}$ is consistent with the SFQM theory. 

\textit{\textcolor{blue}{Nonvolatile electrical control of spin}}--
A central challenge for spin-based devices lies in achieving nonvolatile control without dependence on sustained magnetic fields, gate voltages, or optical excitation. An ideal solution is provided by sliding ferroelectricity/multiferroicity, which facilitates compact, power-efficient, and semiconductor-compatible electrical spin control. This functionality is realized in bilayer Ca(CoN)$_2$ with SFQM, rendering them exceptionally attractive for next-generation applications.

\begin{figure}[htb]
\begin{center}
\includegraphics[width=\columnwidth]{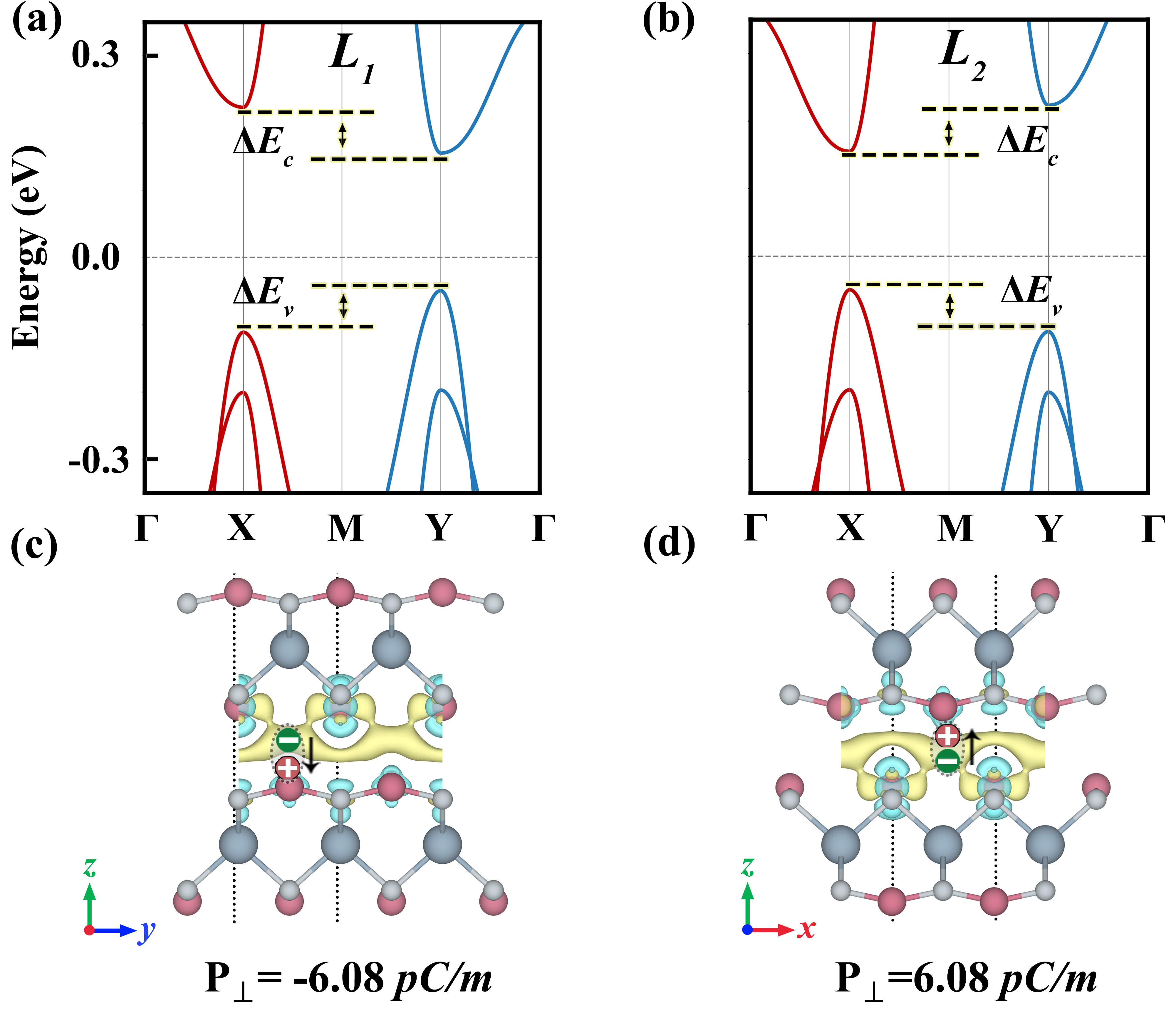}
\caption{(a),(b)  Band structures of  $L_1$, and $L_2$  phase of the bilayer Ca(CoN)$_2$ (without SOC), with red/blue indicating opposite spin polarities. The Fermi level is set to zero. (c),(d) Differential charge density of $L_1$ and $L_2$, where the yellow and cyan regions correspond to charge accumulation and depletion.}
\label{fig:three}
\end{center}
\end{figure}

In the polar-stacked bilayer Ca(CoN)$_2$, the top and bottom layers exhibit identical altermagnetic order.
Consequently, the electronic states $L_1$ and $L_2$ are related by the $\mathcal{O}$ symmetry. This symmetry preserves the spin polarization of the energy bands, as confirmed by the unchanged reverse of the spin-polarized energy band  in the absence of spin-orbit coupling (SOC) in Figs.~\ref{fig:three}{\color{blue}(a)} and~\ref{fig:three}{\color{blue}(b)}. Furthermore, the $H$ phase retains the symmetry operations $\mathcal{O}$ of monolayer Ca(CoN)$_2$ (including $\{C_2||2_{110}| 0\}$, $\{C_2||2_{1-10}| 0\}$, $\{C_2||-4_{001}^{+}| 0\}$, and $\{C_2||-4_{001}^{-}| 0\}$), which gives rise to valley-mediated spin-layer coupling (SLC) due to the presence of valleys at both the $X$ and $Y$ points [Figs.~{\color{blue} S2} and~{\color{blue} S3}].

This SLC allows us to predict the behavior of the spin states in the $L_1$ (or $L_2$) phase. In the conduction band, a nonvolatile electric field (electric polarization $P_{\perp}<0$) pulls down the $X$ valley with spin-up and pushes up the $Y$ valley with spin-down, behaving like a magnetic field pointing along the $z$ direction with $B_{eff}^{c(v)}<0$. In the valence band, the spin splitting is reversed due to the opposite layer polarization [Fig.~{\color{blue} S4}] of the spin states near the $X$ and $Y$ valleys [Fig.~\ref{fig:three}{\color{blue}(b)}], resulting in an effective magnetic field $B_{eff}^{c(v)}>0$. The calculated splitting values are approximately 68 meV at the conduction band minimum (CBM) and 62 meV at the valence band maximum (VBM), significantly exceeding the thermal energy at room temperature ($\sim$26 meV). According to the relation $B_{eff}^{c(v)}=\frac{\Delta E_{c(v)}}{g_s\mu_{B}}$, where $\Delta E_{c(v)}$ is the energy splitting for CBM (VBM), $g_s$ is the effective $g$ factor which here is assumed to be 2, these splitting correspond to an induced magnetic field equivalent to 587 T (535 T).
In contrast, the $L_2$ phase exhibits reversed out-of-plane polarization (electric polarization $P_ \perp>0$), which inverts the layer polarization of spin states near the $X$ and $Y$ valleys and leads to opposite outcomes relative to the $L_1$ phase.

These results align well with the differential charge density distributions in Figs.~\ref{fig:three}{\color{blue}(c)} and~\ref{fig:three}{\color{blue}(d)}. The $L_1$ phase shows a dipole mainly formed by the top layer vacancy and bottom layer Co atom, with charge accumulating around the top layer vacancy and depleting around the bottom layer Co atom. The $L_2$ phase is obtained via $L_2=\mathcal{O}L_1$, which reverses the atomic configuration and fully inverts the out-of-plane polarization. Furthermore, the interlayer sliding in the SFQM mechanism---occurring on the scale of the lattice constant---generates a substantial out-of-plane electric polarization. The $L_1$ ($L_2$) phase exhibits a polarization of $\mp 6.08 \text{ pC/m}$, which is substantially larger than the values reported for other layered systems such as bilayer $h$-BN  ($2.08 \text{ pC/m}$) \cite{hy-li-2017}, bilayer WTe$_2$  ($0.38 \text{ pC/m}$) \cite{hy2-WTe2-bilayer}, bilayer MoS$_2$  ($0.97 ~\text{pC/m}$) \cite{hy-li-2017}, bilayer MnPSe$_3$  ($0.14 \text{ pC/m}$) \cite{MnPSe3}, and bilayer GeSe  ($2.12 \text{ pC/m}$) \cite{GeSe}. This clearly demonstrates the enhanced ferroelectricity in SFQM compared to conventional sliding systems.

\begin{figure}[htb]
	\begin{center}
	\includegraphics[width=\columnwidth]{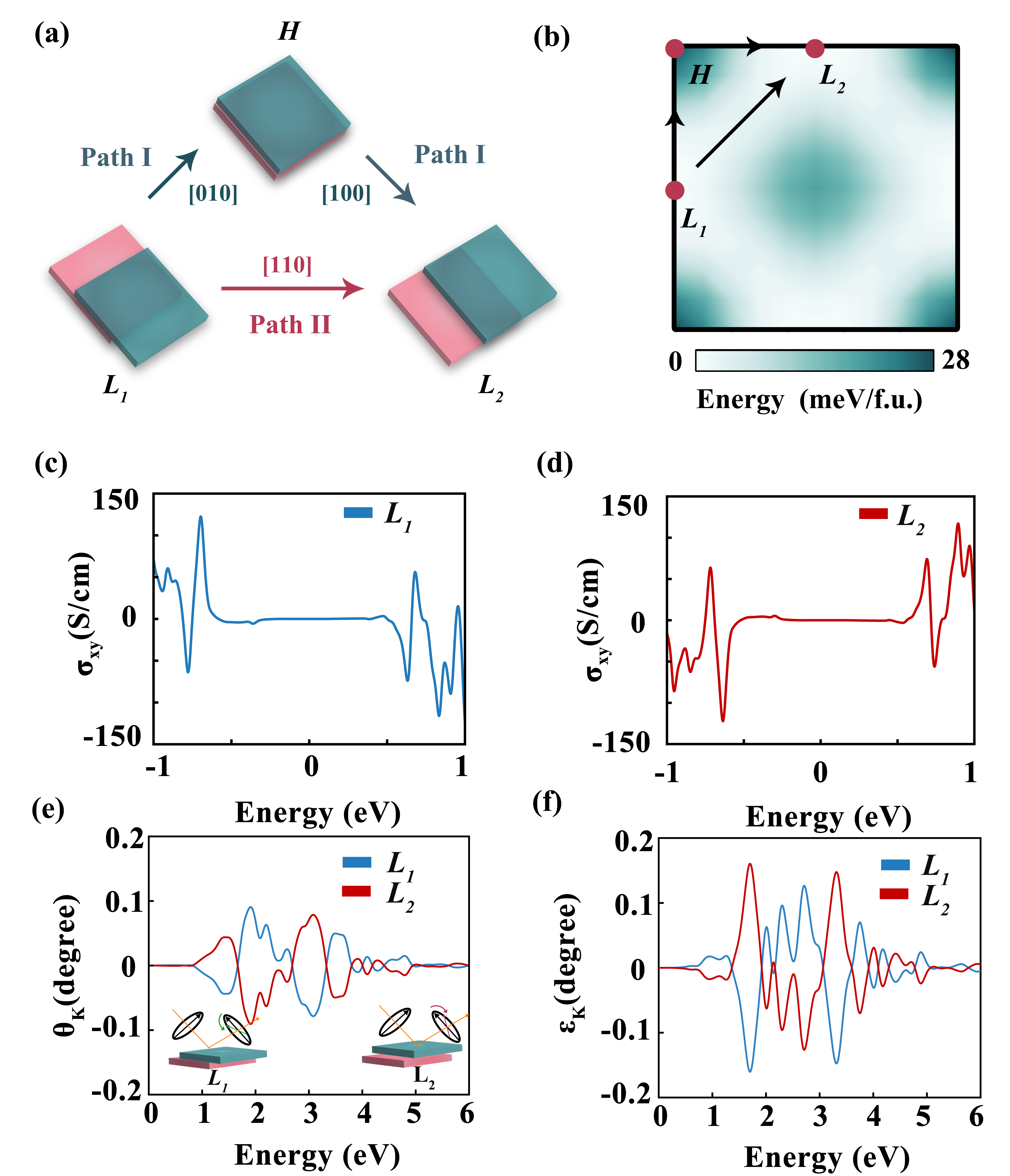}
	\caption{(a) The schematic illustration of two transition  pathways between the $L_1$ and $L_2$ phases. (b) Energy distribution of bilayer Ca(CoN)$_2$ with the ground state energy ($L_{1(2)}$) as the reference. (c),(d) The anomalous Hall conductivities for the $L_1$ phase and $L_2$ phase of bilayer Ca(CoN)$_2$. (e),(f) The magneto-optical response of the $L_1$ and $L_2$ phases for bilayer Ca(CoN)$_2 $, characterized by Kerr angle and Kerr ellipticity.}
	\label{fig:four}
	\end{center}
\end{figure}

\textit{\textcolor{blue}{Discussion}}-- 
The unique sliding pathway in bilayer Ca(CoN)$_2$ enables superlubric switching between the $L_1$ and $L_2$ phases. We identify two distinct transition pathways. In Path-I, the top layer undergoes sequential displacements of $1/2$ unit cell along the [010] and then [100] directions. In contrast, Path-II involves a direct $\sqrt{2}/2$ unit cell shift along the [110] direction [Figs.~\ref{fig:four}{\color{blue}(a)} and~\ref{fig:four}{\color{blue}(b)}]. Notably, Path-II exhibits a significantly lower energy barrier of 2.13 meV/f.u. compared to 27.78 meV/f.u. for Path-I (Fig. S6), establishing it as the energetically favored route. This switching barrier is also orders of magnitude lower than those in typical FQFE materials such as AgBr (155 meV/f.u.), $\alpha$-In$_2$Se$_3$ (68 meV/f.u.), and HgI$_2$ (68 meV/f.u.) \cite{FQFE-NC}, highlighting the potential of SFQM-based devices for high-speed multiferroic memory applications.

The anomalous Hall effect serves as a key signature for identifying SQFM. Interlayer sliding breaks the $\mathcal{O'}$ symmetry, lifting the magnitude degeneracy of $\Omega_z(\mathbf{k})$ near the valleys. This symmetry operation $\mathcal{O'}$, which connects the $L_1$ and $L_2$ phases, inverts the Berry curvature magnitude between the valleys, generating opposite net Berry curvatures (see Fig. {\color{blue} S5} for details). We further calculated the anomalous Hall conductivity [Figs.~\ref{fig:four}{\color{blue}(c)} and~\ref{fig:four}{\color{blue}(d)}] to characterize and contrast the $L_1$ and $L_2$ phases. Together with the layer polarization, these results reveal a layer-locked character and demonstrate that interlayer sliding offers a feasible means to control the electronic states.

The magneto-optical Kerr effect (MOKE) \cite{kerr-soc1,kerr-soc2,kerr-soc3,kerr-soc4} offers a direct means to probe switching between SFQM phases. As shown in Figs.~\ref{fig:four}{\color{blue}(e)} and~\ref{fig:four}{\color{blue}(f)}, MOKE provides a viable method for detecting these phase transitions, particularly where conventional electrical measurements are challenged by the invariant Néel vector during the $L_1$ $\rightarrow$ $L_2$ switching process. Unlike most antiferromagnets that require external fields to break $\mathcal{PT}$ symmetry, SFQM exhibits intrinsic $\mathcal{PT}$-symmetry breaking, enabling direct observation via MOKE. Crucially, the symmetry operation $\mathcal{O'}$ that connects the $L_1$ and $L_2$ phases reverses $\sigma_{xy}$, leading to opposite Kerr rotation angles and ellipticities between the two phases. This sign reversal, combined with the experimental accessibility of MOKE compared to spin-resolved techniques, establishes it as an effective probe for verifying SFQM properties and detecting sliding-induced property inversions.

\textit{\textcolor{blue}{Conclusion}}-- 
In summary, we introduce a new multiferroic phase termed SFQM, in which interlayer sliding drives symmetry reconstruction and enables nonvolatile switching of multiple physical properties. The two intrinsic polar phases, $L_1$ and $L_2$, are related by the symmetry operation 
$\mathcal{O}$ ($\mathcal{O'}$)
, which can be physically realized through interlayer sliding—enabling highly efficient phase control. SFQM uniquely integrates sliding ferroelectricity with fractional quantum ferroelectricity, allowing nonvolatile manipulation of spin (valley) polarization, layer polarization, Berry curvature, and anomalous Hall conductivity. Using bilayer Ca(CoN)$_2$ as a representative system, we demonstrate a novel mechanism for generating fractionally quantized polarization via sliding, establishing a new paradigm in which mechanical sliding replaces conventional electric or magnetic fields for nonvolatile control of electronic states. We further confirm the feasibility of detecting SFQM phases using the MOKE. The design principles presented here are general and can be extended to other van der Waals altermagnetic materials, opening a viable route toward next-generation spintronic devices with nonvolatile functionality and low energy consumption.

\bibliography{ref}



\end{document}